\documentclass{ws-procs975x65}

\def\rmd{\mathrm{d}}
\def\beq{\begin{equation}}
\def\eeq{\end{equation}}

\begin{document}

\title{A Geometrical Approach to Gravitational Lensing Magnification}
\author{Marcus C. Werner$^*$}

\address{Yukawa Institute for Theoretical Physics, Kyoto University,\\
Kyoto, 606-8502, Japan\\
$^*$E-mail: werner@yukawa.kyoto-u.ac.jp}


\begin{abstract}
The standard definition of gravitational lensing magnification is generalized to Lorentzian spacetimes, and it is shown how it can be interpreted geometrically in terms of the van Vleck determinant and the exponential map. This is joint work with Amir B. Aazami (Kavli IPMU, University of Tokyo).
\end{abstract}


\bodymatter


\section{Introduction}
An important observable in gravitational lensing is flux. Suppose $F$ is the observed flux of a lensed image and $F_0$ is the hypothetical flux in the absence of the lens, then the unsigned magnification factor due to gravitational lensing is
\begin{equation}
\mu=\frac{F}{F_0},
\label{magnification}
\end{equation}
which itself is not regarded an observable since $F_0$ is, in general, unknown. Recently, however, the first instance of a strongly lensed type Ia supernova was reported, whose standardizable luminosity makes this also the first direct measurement of $\mu$.\cite{quimby13} As such measurements are becoming more common and allow new tests of the underlying theory, it is also desirable to generalize the standard quasi-Newtonian formulation of $\mu$ (e.g., Ref.~\refcite{schneider92}, p. 161f) to a spacetime setting and better understand its geometrical meaning. In this note, based on Ref.~\refcite{aazami15}, such a definition is discussed in terms of the van Vleck determinant and also the exponential map, which has recently attracted interest in the lensing context.\cite{reimberg13} For the spacetime view of gravitational lensing in general, see e.g. Ref.~\refcite{perlick90}.

\section{Geometrical Background}
\subsection{van Vleck determinant}
In a spacetime defined by a $4$-dimensional Lorentzian manifold $(M,g)$ with signature $(-,+,+,+)$, consider a convex normal neighbourhood $U$ and a chart with $(x^i)\in \mathbb{R}^4$. Let $t\in [0,1]\mapsto \gamma_V(t)$ be a null geodesic segment in $U$ in the direction $\dot{\gamma}_V(0)=V\in T_pM$, and denote the light emission event by $\gamma_V(0)=p$. Since there is a unique geodesic between $p$ and any other $q\in U$, one can define Synge's world function,
\[
\Omega(p,q)=\frac{1}{2}\sigma(p,q)^2,
\]
where the biscalar $\sigma(p,q)$ is the geodesic distance between $p,q \in U$. Note also that $\Omega(p,\gamma_V(t))=0$, of course, but its derivatives need not vanish. Since we will be interested in focusing properties of neighbouring geodesics, it is useful to introduce the Hessian of mixed second derivatives of $\Omega(p,q)$, 
\[
H(p,q)=\det \left[ \frac{\partial^2 \Omega (p,q)}{\partial x^i(p) \partial x^j(q)} \right],
\]
which manifestly depends on the chosen chart. However, this can be turned into a biscalar by means of the metric determinants, and the result
\[
\Delta(p,q)=-\frac{H(p,q)}{\sqrt{\det g(p)\det g(q)}}
\]
is again a biscalar called the van Vleck determinant (see, e.g., Ref.~\refcite{visser93}). For a Minkowski spacetime $(M_0,\eta)$, a direct computation shows that the van Vleck determinant is simply $\Delta_0(p_0,q_0)=1$ for any $p_0,q_0\in M_0$.

\subsection{The exponential map}
Furthermore, since $U$ is assumed to be a convex normal neighbourhood, the exponential map at $p$,
\[
\exp_p: T_pM \rightarrow U, \ V \mapsto q=\exp_p(V)=\gamma_V(1), 
\]
is a diffeomorphism (e.g., Ref.~\refcite{oneill83}, p. 71). This allows the construction of a special chart with respect to an orthonormal basis $\{e_i|_p\}$ of $T_pM$,
\[
X_p: U \rightarrow \mathbb{R}^4, \ X_p(q)=(V^0,\ldots,V^3), \ V=\sum_{i=0}^3 V^ie_i|_p=\exp_p^{-1}(q),
\]
called the normal chart centred at $p$, in which $g(p)=\eta$ and the geodesic is represented by a straight line through the origin,
\[
X_p \circ \gamma_V: t\mapsto (tV^0,\ldots, tV^3). 
\]
We will also use the Jacobian of the exponential map. Recall e.g. from Ref.~\refcite{oneill83}, p. 196, that, in general, given a differentiable map $\phi:M\rightarrow N$ and volume elements $\omega_M, \omega_N$, the Jacobian function is defined by the pullback,
\[
\phi^*(\omega_N)=J(\phi )\omega_M.
\]
Applying this to $\phi=\exp_p$, a direct computation yields
\begin{equation}
J(\exp_p)(q)=\sqrt{-\det g(q)},
\label{jacobian}
\end{equation}
using normal coordinates centred at $p$. This coordinate system can also be used to evaluate the van Vleck determinant, to obtain
\begin{equation}
\Delta(p,q)=\frac{1}{\sqrt{-\det g(q)}},
\label{vanvleck}
\end{equation}
Details can be found in Ref.~\refcite{aazami15}.

\section{Lensing Magnification}
\subsection{Extension to spacetime}
We shall now propose an extension of the gravitational lensing magnification to spacetime. Since the intrinsic luminosity $L$ of an isotropically emitting light source at $p$ and the flux $F$ observed at $q$ are local quantities, the usual definition of the luminosity distance,
\[
D(p,q)=\sqrt{\frac{L(p)}{4\pi F(q)}},
\]
is meaningful in spacetime. Now suppose the light source is represented by a timelike worldline $\gamma(\tau)$ with $\gamma(0)=p$. Using a normal chart in $U$ centred at the emission event $\gamma_V(0)=p$, a classic result by Etherington gives\cite{etherington33}
\begin{equation}
D(p,q)= -\left.\frac{\rmd\Omega (\gamma(\tau),q)}{\rmd \tau}\right|_0\left(-\det g(q)\right)^{\frac{1}{4}}.
\label{etherington}
\end{equation}
Thus, the gravitational lensing magnification (\ref{magnification}) can be recast in terms of the van Vleck determinant using (\ref{vanvleck}) and (\ref{etherington}), to find
\begin{equation}
\mu(p,q)=\frac{D_0(p,q)^2}{D(p,q)^2}=\left(\left.\frac{\frac{\rmd \Omega_0}{\rmd \tau}}{\frac{\rmd \Omega}{\rmd \tau}}\right|_0\right)^2\frac{\Delta}{\Delta_0},
\label{result}
\end{equation}
where the subscript $0$ denotes quantities in a reference spacetime without lens, different from $(M,g)$. Although (\ref{vanvleck}) and (\ref{etherington}) are dependent on the $p$-centred normal chart chosen, the expression for $\mu(p,q)$ contains \textit{only} biscalars, so (\ref{result}) defines unsigned lensing magnification as a chart-independent biscalar in a convex normal neighbourhood $U\subset M$.
\\
In order to make the ratio in (\ref{result}) geometrically meaningful, one has to provide a prescription for comparison. Suppose, for example, that the reference is a Minkowski spacetime $(M_0,\eta)$ obtained as a limit of $M$ for vanishing lens mass. Hence, $\Delta_0=1$ regardless of the $p_0,q_0\in M_0$ considered. Furthermore, a direct computation yields 
\[
\left.\frac{\rmd\Omega (\gamma(\tau),q)}{\rmd \tau}\right|_0=-g(\gamma'(0),V), \ \ \left.\frac{\rmd\Omega_0 (\gamma(\tau),q)}{\rmd \tau}\right|_0=r,
\]
where $r$ is the spatial separation in $M_0$. Assuming this to be fixed, one can define a \textit{spatially-scaled} magnification $\mu_s(p,q)=\mu(p,q)/r^2$. Using (\ref{result}), this quantity may then be written explicitly as a biscalar on $U$,
\begin{equation}
\mu_s(p,q)=\frac{\Delta(p,q)}{g(\gamma'(p),\exp_p^{-1}(q))^2}.
\label{magnification2}
\end{equation}

\subsection{Geometrical interpretation}
These definitions for $\mu$ are seen to be proportional to the van Vleck determinant $\Delta$. Using again $p$-centred normal coordinates in the convex normal neighbourhood $U$, (\ref{jacobian}) and (\ref{vanvleck}) show that $\Delta$ can also be expressed as the inverse of the Jacobian of the exponential map (see, again, Ref.~\refcite{aazami15} for details),  
\[
\Delta(p,q)=\frac{1}{J(\exp_p)(q)}, 
\]
analogous to the standard definition of $\mu$ using the Jacobian determinant of the lensing map\cite{schneider92} Thus, $\mu$ diverges when $J(\exp_p)(q)=0$, that is, when $\exp_p$ ceases to be a diffeomorphism. In this case, $q$ is conjugate to $p$ and there is no longer a unique geodesic between them (c.f. Ref.~\refcite{oneill83}, p. 271, and Ref.~\refcite{aazami15} for details). 

\section{Concluding remarks}
Given a prescription for comparing events with and without lens, (\ref{result}) provides a generalization of the unsigned gravitational lensing magnification to a biscalar in a convex normal neighbourhood of an arbitrary Lorentzian spacetime.
\\
Since the occurrence of multiple regular images of a given light source is important in strong lensing, it will be interesting to see whether a convex normal neighbourhood can be used to describe this, and also how (\ref{result}) may be extended beyond it. Furthermore, this may shed some light on the question whether invariant sums of the \textit{signed} image magnification, which are known so far only for certain classes of lens models in the quasi-Newtonian approximation, exist also in spacetime. 

\section*{Acknowledgments}
This work was supported by the Hakubi Center for Advanced Research, Kyoto University, Kyoto, Japan, and the World Premier International Research Center
Initiative (WPI), MEXT, Japan.

\end{document}